# Near-field thermal transport between twisted bilayer graphene

Fuwei Yang[1,4] and Bai Song[2,3,4*]

[1]*Center for Nano and Micro Mechanics, Tsinghua University, Beijing 100084, China.*

[2]*Department of Energy and Resources Engineering, Peking University, Beijing 100871, China.*

[3]*Department of Advanced Manufacturing and Robotics, Peking University, Beijing 100871, China.*

[4]*Beijing Innovation Center for Engineering Science and Advanced Technology, Peking University, Beijing 100871, China.*

*Corresponding author. Email: songbai@pku.edu.cn

**ABSTRACT:** Active control of heat flow is of both fundamental and applied interest in thermal management and energy conversion. Here, we present a fluctuational electrodynamic study of thermal radiation between twisted bilayer graphene (TBLG), motivated by its unusual and highly tunable plasmonic properties. We show that near-field heat flow can vary by more than 10-fold over only a few degrees of twist, and identify special angles leading to heat flow extrema. These special angles are dictated by the Drude weight in the intraband optical conductivity of TBLG, and are roughly linear with the chemical potential. Further, we observe multiband thermal transport due to the increasing role of interband transitions as the twist angle decreases, in analogy to monolayer graphene in a magnetic field. Our findings are understood via the surface plasmons in TBLG, and highlight its potential for manipulating radiative heat flow.

**INTRODUCTION**

Radiative heat transfer between closely spaced bodies is primarily mediated by the tunneling of evanescent photons[1-3]. Due to the large density of states in the near-field, the energy transfer rate can exceed Stefan-Boltzmann's blackbody limit to far-field radiation by orders of magnitude, especially when various surface modes are populated and coupled[4-10]. Surface modes can also enable quasi-monochromatic thermal radiation, in contrast to the broadband Planck spectrum[11]. With electrically and chemically tunable surface plasmon polaritons (SPPs) in the terahertz and mid-infrared range, graphene offers great potential for enhancing and manipulating near-field radiative heat transfer (NFRHT) over a wide temperature range[12-26]. To this end, both suspended and supported monolayer graphene have been extensively analyzed. Further, graphene SPPs also offer rich opportunities for harnessing thermal radiation in diverse energy conversion and heat management devices including thermophotovoltaic cells[27, 28], thermal switches[29, 30], and rectifiers[31, 32].

The magic of monolayer graphene (MLG) in NFRHT ultimately lies in its tunable optical conductivity and non-local dielectric function, which originate from its characteristic Dirac-cone electronic band structure[33]. Recently, bilayer graphene has drawn tremendous attention because a small twist angle between the two layers lead to a variety of Moiré patterns, which can substantially alter the band structure and reveal a range of exotic phenomena including correlated insulators and unconventional superconductivity[34-36]. The low-energy band structure of twisted bilayer graphene (TBLG) can be approximately viewed as two shifted MLG Dirac cones, with the Moiré

Brillouin zone determined by the shift vector (Fig. 1). In addition to a strong optical response similar to MLG, TBLG features a complex low-energy spectrum with multiple interband peaks that are particularly sensitive to the twist angle[37-42]. Despite its rich and unusual electronic and optical properties, the characteristics and potentials of TBLG in NFRHT remain to be explored.

Here, we theoretically study near-field thermal transport between two suspended TBLG sheets separated by a vacuum gap within the framework of fluctuational electrodynamics (Fig. 1). We employ the effective continuum model and linear response theory to calculate the electronic structure and optical response of TBLG, respectively, and show that the photon-mediated heat flow can be substantially controlled via the twist angle. In particular, a small twist around some chemical-potential-dependent special angles can lead to over 10-fold heat transfer enhancement. This dramatic variation is understood via the Drude weight—a key parameter characterizing the intraband optical response of TBLG. In addition, we find that as the twist angle decreases interband transitions become increasingly important, leading to multiband thermal transport similar to MLG in a magnetic field. We conclude with a discussion of the essential role of SPPs in the NFRHT of TBLG.

**METHODS**

**Near-field Thermal Transport.** Based on the theoretical framework of fluctuational electrodynamics, the total heat flux between two parallel planes across a vacuum gap $d$ (Fig. 1a) can be expressed in the Landauer form as[3]

$$q(T_1, T_2, d) = \int_0^\infty d\omega [\Theta(\omega, T_1) - \Theta(\omega, T_2)] f(\omega). \tag{1}$$

Here, $\Theta(\omega, T) = \hbar\omega/[\exp(\hbar\omega/k_B T) - 1]$ is the mean energy of a harmonic oscillator less the zero-point contribution, and $f(\omega) = \int_0^\infty dk \frac{k}{4\pi^2} [\tau_s(\omega, k) + \tau_p(\omega, k)]$ is the spectral transfer function, with $T$, $\omega$, and $k$ being the temperature, frequency, and wavevector component parallel to the planes, respectively. $\tau_s$ and $\tau_p$ are the photon tunneling probabilities for the *s*- and *p*-polarized waves given by

$$\tau_{\alpha=s,p}^{12}(\omega, k) = \begin{cases} \dfrac{(1 - |r_\alpha^1|^2 - |t_\alpha^1|^2)(1 - |r_\alpha^2|^2 - |t_\alpha^2|^2)}{|D_\alpha|^2}, & \text{if } k < \omega/c \\ \dfrac{4\mathrm{Im}(r_\alpha^1)\mathrm{Im}(r_\alpha^2) e^{-2\mathrm{Im}(\zeta)d}}{|D_\alpha|^2}, & \text{if } k > \omega/c \end{cases}, \tag{2}$$

where $r_\alpha^i$ and $t_\alpha^i$ are the Fresnel coefficients which for graphene are often written as functions of the optical conductivity (see supporting information), $D_\alpha = 1 - r_\alpha^1 r_\alpha^2 e^{2i\zeta d}$ is a Fabre-Pérot-like denominator, and $\zeta$ is the transverse wavevector component. We further define the spectral and total heat transfer coefficient in the linear regime as $h_\omega = \frac{\partial \Theta(\omega,T)}{\partial T} f(\omega)$ and $h = \int_0^\infty d\omega h_\omega$, respectively.

**Band Structure of TBLG.** To obtain the optical conductivity of TBLG, we start by calculating its electronic band structure. We focus on the regime of low energy ($\leq 1$ eV) and small twist angles (~2° to 8°), which is well described by the widely used effective continuum model[34, 39, 42-44]. The Hamiltonian is usually written as

$$H = \begin{pmatrix} H_1 & U \\ U^\dagger & H_2 \end{pmatrix}, \tag{3}$$

where $H_l$ ($l$ = 1, 2) is the Dirac-Hamiltonian for the two layers, while $U$ denotes interlayer hopping. Expanding around K$_1$ and K$_2$ (Fig. 1b), $H_l$ can be written as

$$H_l(\boldsymbol{k}) = \hbar v_F (\boldsymbol{k} - \boldsymbol{K}_l) \cdot \boldsymbol{\sigma}. \tag{4}$$

Here, $\boldsymbol{\sigma}$ denotes the Pauli matrices, $v_F = \sqrt{3} a_0 t_0 / 2\hbar \approx 10^6 \, m/s$ is the Fermi velocity, with the graphene lattice constant $a_0 = 2.46$ Å and the intralayer hopping integral $t_0$ chosen as 2.78 eV. Considering nearest-neighbor coupling only, the interlayer hopping term is given by[34]

$$U = t_\perp \sum_{j=1}^{3} \exp(-i\boldsymbol{q}_j \cdot \boldsymbol{r}) U^j, \qquad (5)$$

with $U^1 = \begin{pmatrix} 1 & 1 \\ 1 & 1 \end{pmatrix}$, $U^2 = \begin{pmatrix} e^{i\phi} & 1 \\ e^{-i\phi} & e^{i\phi} \end{pmatrix}$, $U^3 = \begin{pmatrix} e^{-i\phi} & 1 \\ e^{i\phi} & e^{-i\phi} \end{pmatrix}$, and $\phi = \frac{2}{3}\pi$. $\boldsymbol{q}_j$ represents interlayer hopping (Fig. 1b) with an energy of $t_\perp = 0.11$ eV.

As an example, the calculated band structure for TBLG with a twist angle of 3° is shown in Fig. 2a. In the vicinity of the K (K') point, the typical linear dispersion for monolayer graphene remains. However, the band structure becomes rather complex with increasing energy. In particular, the appearance of saddle points due to anti-crossing at the band intersections of different layers leads to van Hove singularities in the density of states[40], which substantially affect optical transitions in TBLG.

**Optical Conductivity of TBLG.** We first compute the real part of the optical conductivity ($\sigma$) of TBLG from its band structure via the Kubo formula[39, 42]. Subsequently, the imaginary part is obtained through the Kramers-Kronig relation[42]. For graphene, $\sigma$ is often divided into an intraband (Drude) and an interband term as

$$\sigma(\omega) = \sigma_D + \sigma_I. \qquad (6)$$

The intraband conductivity can be written as

$$\sigma_D(\omega) = \frac{D}{\pi} \frac{i}{\hbar\omega + i\Gamma}, \qquad (7)$$

where $\Gamma$ represents the electron scattering rate and is chosen as 7 meV, $D$ is the Drude weight which characterizes the strength of intraband transitions[33] and is calculated as[42]

$$D = \frac{4\sigma_0}{\pi\hbar} \int d\boldsymbol{k} \sum_m \left(\frac{\partial \epsilon_{m,\boldsymbol{k}}}{\partial k_x}\right)^2 \frac{\partial n_F(\epsilon_{m,\boldsymbol{k}})}{\partial \epsilon}. \qquad (8)$$

Here, $n_F(\epsilon) = 1/[\exp[(\epsilon - \mu)/k_B T] - 1]$ is the Fermi distribution function, $\mu$ is the chemical potential, $\epsilon_{m,k}$ is the energy of the $m^{th}$ band with momentum $k$, and $\sigma_0 = e^2/4\hbar$. The integration is taken over the first Moiré Brillouin zone, and only the x-component is considered because of the hexagonal symmetry in TBLG[42]. For the interband conductivity, the real part is calculated as[39]

$$\mathrm{Re}\,\sigma_I(\omega) = \frac{4\sigma_0}{\pi} \int d^2k \sum_{mn} [n_F(\epsilon_{n,k}) - n_F(\epsilon_{m,k})] |\langle m,k|j_x|n,k\rangle|^2 \frac{\delta(\hbar\omega - \epsilon_{m,k} + \epsilon_{n,k})}{\epsilon_{m,k} - \epsilon_{n,k}}. \quad (9)$$

Here, $j_x = -\frac{\partial H}{\partial k_x}$ is the current operator, $|n,k\rangle$ represents the eigenstates, and $\delta(x)$ is replaced by $\Gamma/\pi/(x^2 + \Gamma^2)$ in numerical computation.

In Fig. 2b, we show the calculated optical conductivity at $\mu$ = 0.25 eV for MLG, decoupled bilayer graphene (DBLG), and TBLG with $\theta$ = 2°, 3° and 8°. Ignoring interlayer coupling, the optical conductivity of DBLG is simply twice that of MLG[42]. At a large twist angle of 8°, the optical conductivity of TBLG approaches that of DBLG with minor differences at relatively high energy. As the twist angle decreases, a series of low-energy peaks appear due to interband transitions, in clear contrast to DBLG. After obtaining the real part of the interband conductivity, the imaginary part is then calculated as[42]

$$\mathrm{Im}\,\sigma_I(\omega) = \frac{2\omega}{\pi} \mathcal{P} \int_0^\infty dv \frac{\mathrm{Re}\sigma_I(v) - 2\sigma_0}{\omega^2 - v^2}, \quad (10)$$

where $\mathcal{P}$ denotes Cauchy principal integral.

## RESULTS AND DISCUSSION

**Twist-Angle-Sensitive Heat Flow.** The significant effect of the twist angle on NFRHT is firstly manifested in the total heat transfer coefficient (HTC). In Fig. 3a, we plot the HTC for MLG, DBLG, and TBLG with $\theta$ = 8°, 3°, and 2°, at $T$ = 300 K, $\mu$ = 0.25 eV, and gap sizes from 100 μm to 1 nm. All cases show a dramatic HTC enhancement with reducing gap, eventually exceeding the blackbody limit by over three orders of magnitude. The HTC of TBLG with an 8° twist ($h_{\theta=8°}$) overlaps with that of DBLG ($h_{DBLG}$), since their optical conductivities converge at very low energy (Fig. 2b). At nanometer gaps, $h_{\theta=8°}$ is roughly half of $h_{MLG}$. As the twist angle reduces to 3° and further to 2°, the HTC first increases to $\sim 7h_{MLG}$ and then drops back to $\sim 3h_{MLG}$. At micrometer gaps, however, the opposite trend is observed, with $h_{\theta=8°}$ being the largest and $h_{\theta=3°}$ the smallest.

In order to understand the twist-induced HTC suppression and enhancement, we performed systematic calculations for $\theta$ from 1.5° to 8° at two representative gap sizes of $d$ = 10 nm and 1 μm (Fig. 3b). At small twist angles, $h_{TBLG}$ clearly forms a peak for the 10 nm gap. Interestingly, for the 1 μm gap, a dip appears instead at about the same twist angle, which is roughly 3° for $\mu$ = 0.25 eV. This contrast between small and large gaps is discussed later via the coupling of SPPs in TBLG. As the twist angle increases beyond ~6°, $h_{TBLG}$ consistently approaches $h_{DBLG}$ regardless of the gap size. For even larger $\theta$ which can be treated with a tight-binding model[40], we omit an exact calculation but argue that $h_{TBLG}$ should remain close to $h_{DBLG}$ since the optical conductivities only differ at very high energy which has little contribution to heat transfer.

**Special Angle Identification via the Drude Weight.** Our results above demonstrate that it is possible to tune the near-field radiative heat flow by over 10-fold with only a few degrees of twist around some special angle. In Fig. 3b, we show that at $\mu = 0.25$ eV, the special angle for the maximum (10 nm gap) and minimum (1 μm gap) of $h_{TBLG}$ is also where the Drude weight reaches its first minimum as the twist angle decreases. Such observation holds at other chemical potentials (see Fig. S1 for $\mu = 0.15$ eV and 0.35 eV) as well, suggesting that the Drude weight can be used to quickly identify special angles for heat flow extrema without actually computing NFRHT, which demands the resource-intensive computation of the full conductivity. The reason that the Drude weight can be a good indicator of the heat flow is twofold. First, the Drude conductivity dominates at low energy except for twist angles about or below 2°. Secondly, only low-energy optical responses matter in NFRHT near room temperature.

In Fig. 3c, we further plot the Drude weight as a function of $\theta$ for a range of chemical potentials from 0.15 eV to 0.45 eV. By extracting the largest twist angle where the Drude weight is a minimum in each case, we observe an approximately linear relation between the special angle and the chemical potential (Fig. 3d), which offers a convenient way to identify these angles. Despite the observation of an approximate correspondence between the Drude weight and the heat flow at relatively large angles, we also notice a clear deviation at small gaps for twist angles $\lesssim 2°$ (Fig. 3b), which signifies an increasing influence of the interband transitions.

**Multiband Transport at Small Twist Angles.** Motivated by the appearance of multiple low-energy peaks in the optical conductivity with decreasing twist angle (Fig. 2b), we now explore the spectral characteristics of NFRHT in TBLG. To this end, we compare TBLG with $\theta = 2°$ to DBLG at $\mu = 0$ eV, $d = 10$ nm, and $T = 300$ K, 600 K, and 1000 K, without considering the temperature dependence of optical conductivity. The spectral heat transfer coefficient of DBLG (Fig. 4a) features a single peak at all three temperatures. The spectral HTC of TBLG also shows a single peak (with a small shoulder) at room temperature, however, as temperature increases, an increasing number of peaks become visible. These multiband transport spectra apparently arise from the interband transitions, with the temperature serving as an energy filter through the Boltzmann factor $\Theta(\omega, T)$.

To further explore the underlying mechanism, we focus on the spectral transfer function so that the temperature factor is dropped. We first note that $f_{TBLG}$ almost overlaps with $f_{DBLG}$ at the lowest energies where Drude conductivity dominates. More interestingly, $f_{TBLG}$ ($\theta = 2°$) exhibits multiple peaks and dips absent in $f_{DBLG}$, whose positions coincide with those in the imaginary part of the optical conductivity (Fig. 4c). Specifically, $f_{TBLG}$ drops sharply whenever Im $\sigma$ becomes negative which leads to optical gaps where SPPs can no longer be sustained[33, 39]. Similar multiband transport was also predicted for MLG in a strong magnetic field[45, 46], which was explained by oscillations of Im $\sigma$ around 0 due to interband transitions between the Landau levels. The above results suggest that TBLG can also be used to control the radiation spectrum.

**Surface Plasmon Polaritons in TBLG.** To gain a deeper insight into the unique characteristics of TBLG in NFRHT, we analyze the photon tunneling probability and highlight the contributions of *p*-polarized surface plasmon polaritons. In Fig. 5, we present $\tau_p$ for MLG and TBLG with $\theta = 8°$, $3°$, and $2°$ at $\mu = 0.25$ eV and $d = 10$ nm. In addition, we calculate the dispersion relations of the coupled SPPs by setting $D_p = 1 - r_p^1 r_p^2 e^{2i\zeta d} = 0$, and plot them on top of the $\tau_p$ maps. The overlap between the dispersion curves and the peaks of $\tau_p$ confirms that the coupled SPPs are the dominant mediator for the large near-field HTC of TBLG[7]. For $\theta = 8°$, the dispersion curve (Fig. 5b) is similar to that of MLG (Fig. 5a), with well-defined acoustic and optical branches merging at sufficiently large wavevectors[33]. Compared to $h_{MLG}$, the smaller $h_{\theta=8°}$ (Fig. 3) is due to a shift of the mode-merging region to higher energy and lower wavevector. We note that this shift results from an increase in the Drude weight, since the slope of the intraband plasmon dispersion scales as $\sqrt{D}$[33, 39].

At the smaller twist angles of 3° and 2° (Fig. 5c, d), the dispersions noticeably deviate from that of MLG, as a result of a growing influence of interband transitions on the intraband plasmons, which explains the lack of correspondence between the Drude weight and the HTC at small angles (Fig. 3b). In addition, multiple high-transmission regions appear which are absent in MLG, and shift to lower energy as $\theta$ decreases. These regions represent contributions from the unconventional interband plasmons described in previous work[39], which together with the gaps in-between result in the multiband transport in TBLG.

In general, the above discussions are valid for nanometer gaps. As the gap size increases, the attenuation of SPPs limits any effective coupling to smaller wavevectors where intraband transitions become even more dominant. For intraband plasmons, the attenuation length is proportional to the Drude weight[33]. Therefore, a larger Drude weight leads to stronger coupling and thus higher HTC at sufficiently large gaps, contrary to the case of small gaps (Fig. 3b).

**CONCLUSION**

In summary, we theoretically demonstrate that near-field radiative thermal transport between two suspended twisted bilayer graphene sheets can be effectively controlled by the twist angle. In terms of heat flow, we observe over 10-fold variation as the twist angle varies by only a few degrees around some special angle, which are dictated by the Drude weight that characterizes the intraband plasmons of TBLG, and scales approximately linearly with the chemical potential. In addition, multiband thermal transport occurs at sufficiently small twist angles, as the interband plasmons become increasingly important. Our current investigation is limited to chemical potentials below 0.45 eV and twist angles from 1.5° to 8°. For smaller twist angles approaching the first magic angle, lattice relaxation becomes non-negligible[47-49], which represents the topic of an ongoing study. The effects of transport temperature and the electron scattering rate are also being systematically explored. Our results offer a new way of controlling radiative heat flow, and open yet another avenue for uncovering the full potential of the twist degree of freedom in graphene and similar materials.

ACKNOWLEDGEMENT

This work was supported by the National Natural Science Foundation of China (Grant No. 52076002), the Beijing Innovation Center for Engineering Science and Advanced Technology, the Xplorer Prize from the Tencent Foundation, and the Tsien Excellence in Engineering program. We thank Qizhang Li and Haiyu He for helpful discussions.

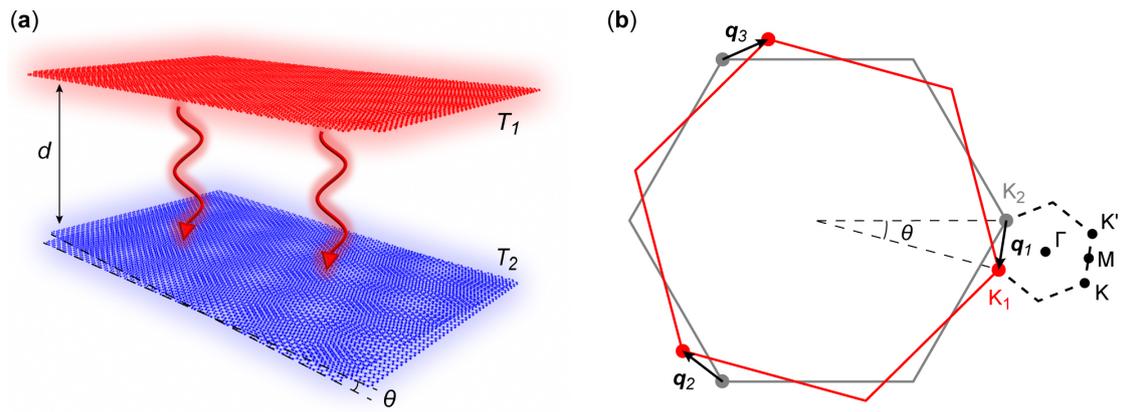

**Figure 1.** (a) Schematic of radiative thermal transport between two identical TBLG sheets separated by a vacuum gap. (b) Reciprocal space of TBLG. The red and gray hexagons represent the first Brillouin zones for the two layers. $K_1$ and $K_2$ are the Dirac points. The dashed black hexagon shows the Moiré Brillouin zone with the high symmetry points labeled. $\boldsymbol{q}_j$ represents the nearest-neighbor interlayer hopping process.

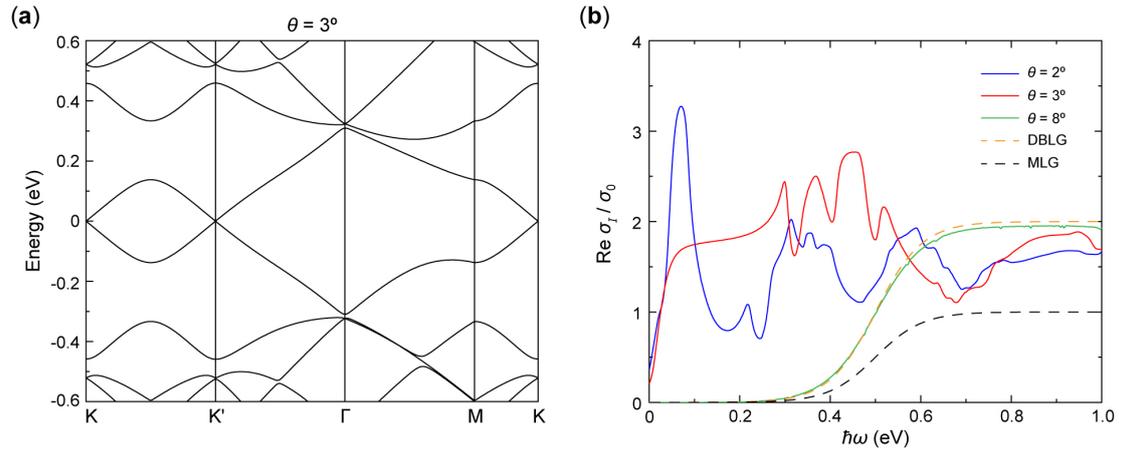

**Figure 2.** (a) The calculated electronic band structure of TBLG with a twist angle of 3°. (b) Real part of the interband optical conductivity of MLG, DBLG, and TBLG with representative twist angles. $\sigma_0 = e^2/4\hbar$ is the universal conductivity of MLG.

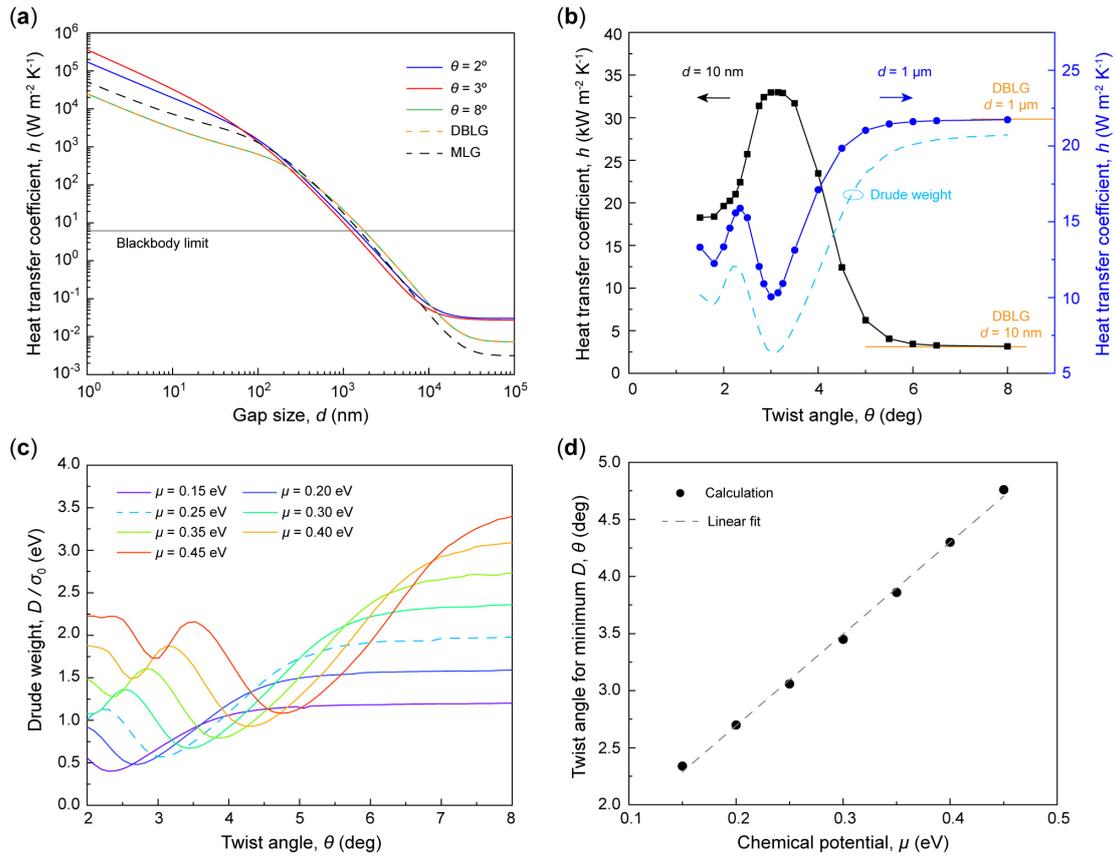

**Figure 3. Twist-angle-dependent NFRHT.** (a) Heat transfer coefficient as a function of gap size for representative cases at $T$ = 300 K and $\mu$ = 0.25 eV. (b) Heat transfer coefficient as a function of twist angle for $d$ = 10 nm and $d$ = 1 μm. The dashed cyan curve shows the corresponding Drude weight. (c) Variation of Drude weight with twist angle at different chemical potentials, and (d) the respective twist angle for minimum Drude weight. The gray dashed line is a linear fit.

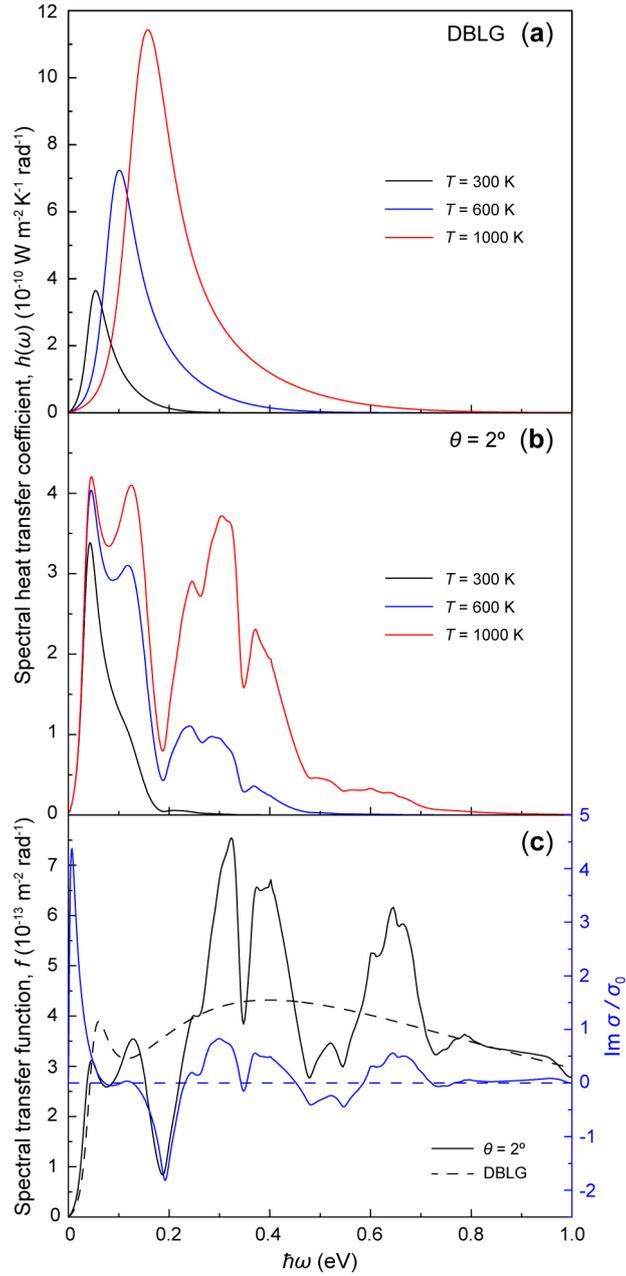

**Figure 4. Multiband transport in TBLG.** Spectral heat transfer coefficient for (a) DBLG and (b) TBLG with $\theta = 2°$ at room and higher temperatures, with $\mu = 0$ eV and $d = 10$ nm. The spectral transfer function (left) for DBLG and TBLG with $\theta = 2°$, and the imaginary part of the optical conductivity of TBLG (right). The dashed blue line indicates Im $\sigma = 0$.

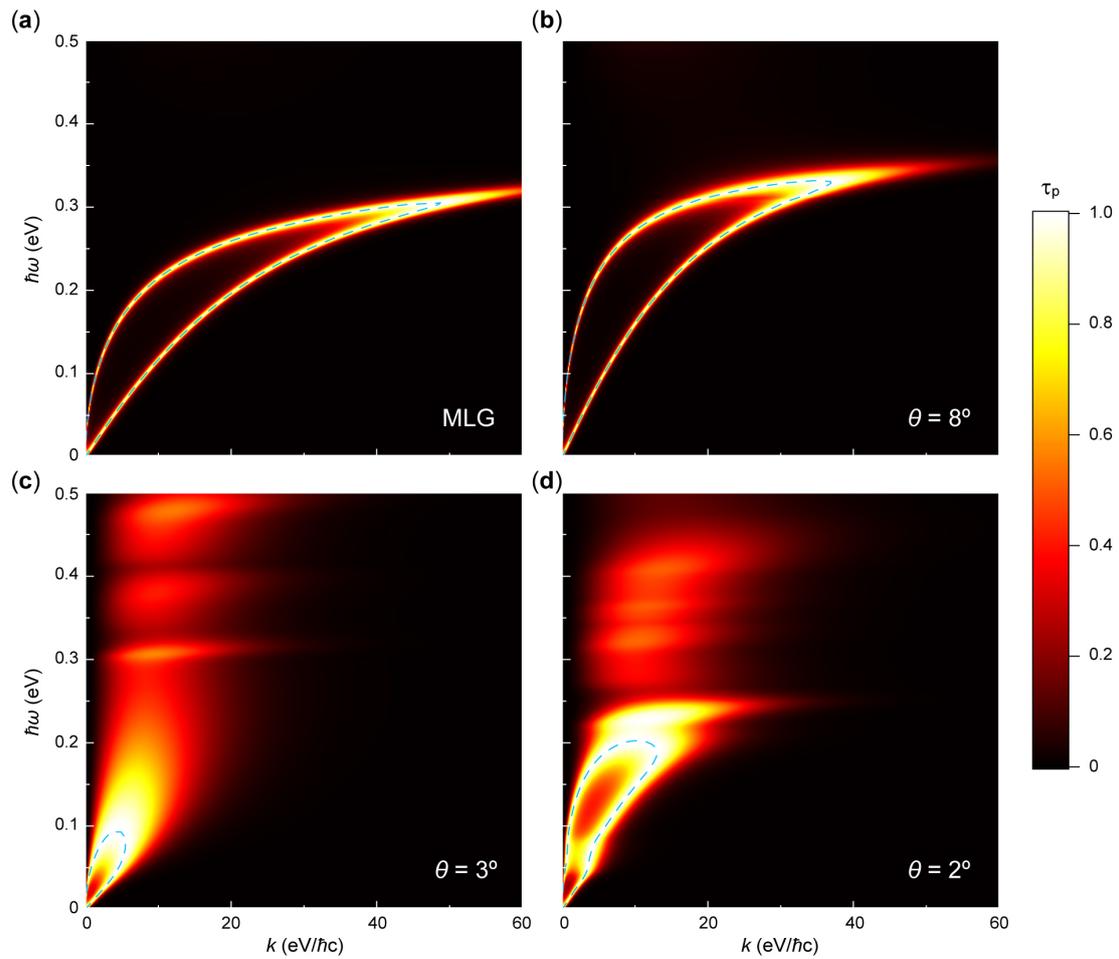

**Figure 5. Photon tunneling probability across a 10 nm gap at $\mu = 0.25$ eV.** (a) MLG. (b)-(d) TBLG with $\theta = 8°$, $3°$, and $2°$, respectively. The dashed cyan lines represent the dispersions of coupled SPPs.